\documentclass[preprint,3p,12pt]{elsarticle}

\journal{Superlattices and Microstructures}
\usepackage{color}
\usepackage{graphicx,subfigure}
\usepackage{longtable}
\usepackage{textcomp}
\usepackage{amsmath}

\begin{document}

\begin{frontmatter}

\title{Structural and Electronic properties of cubic (GaN)$_1$/(ZnO)$_1$ superlattice: \\
Modified Becke-Johnson exchange potential}

\author[univ]{M.R. BOUFATAH}
\ead{boufatah@gmail.com}

\author[univ,ictp]{A.E. MERAD\corref{cor1}}
\ead{merad{\_}a@yahoo.fr}

\address[univ]{Laboratoire de Physique Th\'eorique, D\'epartement de Physique, 
         Facult\'e des Sciences, Universit\'e de Tlemcen, 
         B.P. 119 Tlemcen 13000 Alg\'erie}
\address[ictp]{The Abdus Salam ICTP, Strada Costiera, 11 I-34014 Trieste, Italy}

\cortext[cor1]{Corresponding author}

\begin{abstract}

The structural and electronic properties of new structural cubic (GaN)$_1$/(ZnO)$_1$ superlattice have been investigated using two different theoretical techniques: the full potential-linearized augmented plane wave (FP-LAPW) method and the linear combination of localized pseudo atomic orbital (LCPAO). The new modified Becke-Johnson (mBJ) exchange potential is chosen to improve the bandgap of the superlattice and effective masses. The bandgap is found to be slightly indirect and reduced from those of pure GaN and ZnO. The origin of this reduction is attributed to the $p-d$ repulsion of the Zn-N interface and the presence of the O $p$ electron. The electron effective mass is found to be isotropic. Good agreement is obtained between two used methods and with available theoretical and experimental data.

\end{abstract}


\begin{keyword}
 Ab initio \sep FP-LAPW \sep LCPAO \sep Oxynitride \sep GaN/ZnO superlattice \sep Band structure \sep Density of states \sep Effective mass \sep mBJ-(LDA,GGA)
\PACS 71.15.-m \sep 71.18.+y \sep 71.20.-b \sep 71.20.Mq \sep 71.20.Nr \sep 81.05.Dz \sep 81.05.Ea
\end{keyword}


\end{frontmatter}


\section{Introduction}
\label{intro} 
The (GaN)$_1$/(ZnO)$_1$ pseudobinary semiconductor alloys, which is discovered in this last years \cite{maeda2005,maeda2006} represents a new class of alloys typed as III-V/II-VI. This latter is different from classical isovalent semiconductors alloys (i.e, III-V/III-V, II-VI/II-VI). From technological aspect, the growth of ZnO on GaN heterostructure was realized using the vapor cooling condensation system and the related heterojunctions light-emitting diodes LEDs are then fabricated \cite{ricky07}. But the main technological application is the hydrogen generation from water photosplitting through the photo-electrochemical cell (PEC). In fact, under visible light, it has a much higher water splitting efficiency compared to other oxides \cite{lewis07}. The spectacular feature of this characteristic is attributed partly to the lower bandgap of alloys and in particular of the heterostructure compared to its parent elements GaN and ZnO. Few theoretical and experimental studies estimating the optical bowing parameter have been reported \cite{maeda2005c,huda08,wang2010}. Large difference in the estimation values was than observed. The recent calculation of S. Wang and L-W Wang \cite{wang2010} shows that the ordering plays an important role in the reduction of the bowing parameter.\\
In the other hand, the tendency in this decade of several research groups to elaborate the GaN and ZnO in the zinc-blend structure for technological interest, motivate us to contribute by a predictive study for this (GaN)$_1$/(ZnO)$_1$ superlattice in this structural phase. In fact the lattice constants of GaN and ZnO are nearly identical and therefore this superlattice in the zinc-blend structure is then possible to growth. Regarding the electronic properties from theoretical point of view, the problem of the underestimation of the bandgap using the local density and generalized gradient approximations (LDA and GGA) of density functional theory (DFT) is very remarkably for ZnO in particular. The recent works have predicted the values of 0.710 eV \cite{zhu08} for ZnO and 1.811 eV \cite{li11} for GaN which are lowered from the experimental ones by $78 \%$ and $45 \%$ respectively. 
Consequently, this error affects considerably the value of the large bandgap discontinuity at the heterointerface for (GaN)$_1$/(ZnO)$_1$ layered structure. In order to overcome this problem and improve the bandgap of the superlattice we used the most recent proposed approximation called modified Becke-Johnson (mBJ) exchange potential \cite{koller11}. In fact, this technique is capable to describe with high accuracy the electronic structure of semiconductors and insulators giving rise to a significantly improved bandgap values to be much closer to the experimental ones. Because of the lake of data about the structural and electronic properties for the studied superlattice we used two different ab initio techniques FP-LAPW and LCPAO in order to help understanding the related properties and consolidate our results.\\
The paper is organized as follows: In section \ref{compute}, we describe briefly the computational techniques used in this study. In section \ref{results}, we present and discuss our obtained results for the electronic properties of the studied superlattice. We summarize our main conclusions in section \ref{conclusion}.

\section{Computational details}
\label{compute}

\begin{figure}[h]
\begin{center}
\includegraphics[width=12cm]{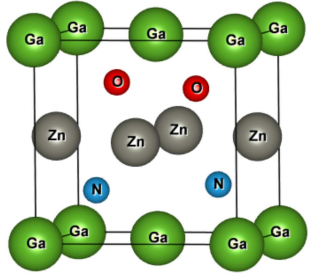} 
\caption{The $1 \times 1 \times 1$ supercell of (GaN)$_1$/(ZnO)$_1$ (atomic color map: green for Ga, blue for N, grey for Zn and red for O)}
\label{crystal}
\end{center}
\end{figure}

The simulation of the (GaN)$_1$/(ZnO)$_1$ structure in the (001) direction is achieved with cubic $1 \times 1 \times 1$ supercell containing 8 atoms where the atomic positions are those of cubic zinc-blende (Figure \ref{crystal}). The calculations were performed using both full potential-linearized augmented plane wave (FP-LAPW) and linear combination of localized pseudo atomic orbital (LCPAO) methods, based on the density functional theory (DFT) \cite{kohn64,dreizler90}. 
The description of the specific parameters to each technique is given as :

\subsection{FP-LAPW}
\label{computeLAPW}

We used the full-potential linear augmented plane-wave method (FP-LAPW) implemented in the Wien2k code \cite{blaha01} which self-consistently, finds the eingenvalues and eingenfunction of the Kohn-Sham equations for the system \cite{kohn64,dreizler90}. We used the Generalised Gradient Approximation (GGA) as parameterized by Perdew, Burke and Ernzenhorf \cite{PBE96a}, which includes the second order gradient components, and Local Density Approximation (LDA) both with modified Becke-Johnson exchange potential (mBJ). The core states of Ga, N, Zn and O atoms are self-consistently treated and relativistically-relaxed in a spherical approximation, whereas the valence states are treated self-consistently within the semi-relativistic approximation (spin–orbit coupling excluded). The valence electron configurations used in the calculations are: Ga$(3d^{10}4s^24p^1)$, N$(2s^22p^3)$, Zn$(3d^{10}4s^2)$ and O$(2s^22p^4)$. The wave function, charge density and potential are expanded by spherical harmonic functions inside non-overlapping spheres surrounding the atomic sites (muffin-tin spheres) and by a plane-wave basis set in the remaining space of the unit cell (interstitial region). The maximum $l$ quantum number for the wave function expansion inside atomic spheres is confined to $l_{max}=10$. The charge density is Fourier-expanded up to $G_{max}=8.5 ~(Ry)^{1/2}$. The convergence parameter $R_{MT}k_{max}$ which controls the size of the basis set in these calculations is set to $7$ ($k_{max}$ is the maximum modulus for the reciprocal lattice vector and $R_{MT}$ is the average radius of the muffin tin spheres). The muffin tin radius values for Ga, N, Zn and O were chosen at $1.96$, $1.68$, $2$ and $1.73$ atomic units (a.u.) respectively. The reciprocal space is sampled by a $12 \times 12 \times 12$ Monkhorst-Pack mesh \cite{monkhost76} with $868k$-vectors in the irreducible Brillouin zone. The iteration process is repeated until the calculated total energy of the crystal converges to less than $10^{-5} ~Ry$.

\subsection{LCPAO}
\label{computeLCPAO}

For LCPAO method \cite{ozaki05a}, we used $6\times6\times6$ $k$ points for the $k$-space integration. The primitive pseudo atomic basis sets Ga7.0-$s3p3d3f1$, Zn6.0-$s3p3d3f1$, N6.0-$s3p3d2$ and O7.0-$s3p3d3$ are used for Ga, Zn, N and O atoms, respectively, where the abbreviation, for exemple, Ga7.0-$s3p3d3f1$, represents the employement of three primitive $s$ orbitals, three primitive $p$ orbitals, three primitive $d$ orbitals and one primitive $f$ orbitals of a Ga atom which are generated with a confinement radius of $7.0 ~bohr$. The standard DFT PP for Zn atom treat $3d^{10}$, $4s^2$ as the valence electrons. In this work, we include the $3s$ and $3p$ states as the valence electrons to improve accurate calculations \cite{dixit2011}. The pseudopotentiels used in the present calculations were avaible on the openmx web site \cite{openmx}. The generalized gradient functional developed by Perdew et al \cite{PBE96a} is used to evaluate the exchange-correlation energy. The real space grid techniques are used with the cutoff energy of $450 ~Ry$ in numerical integrations and solution of the Poisson equation using the fast Fourier transformation (FFT) algorithm. The equilibrium positions of ions are reached by the the eigenvector following method where the approximate Hessian is updated by the Broyden-Fletcher-Goldfarb-Shanno (BFGS) method \cite{openmx}. The Hellmann-Feynman forces tolerence is $1 ~mRy/bohr$. All total energies were converged to $2.0 \times 10^{-10}~Ry$.

\section{Results and discussion}
  \label{results}

\subsection{Structural properties}
  \label{structural}

\begin{figure}[!htp]
 \centering
 \subfigure[]{\label{optim_LDA-LAPW} \includegraphics[width=7.00cm]{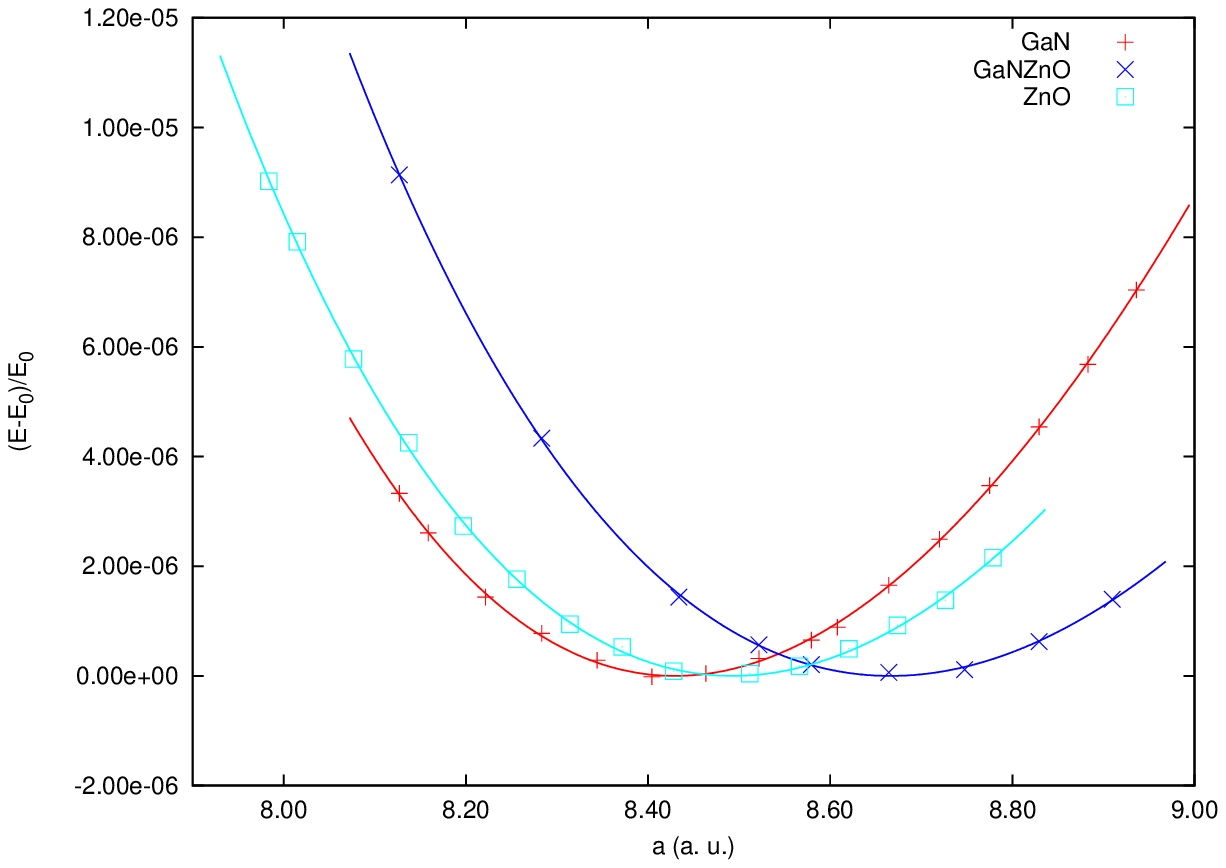}}
 \subfigure[]{\label{optim_PBE-LAPW} \includegraphics[width=7.00cm]{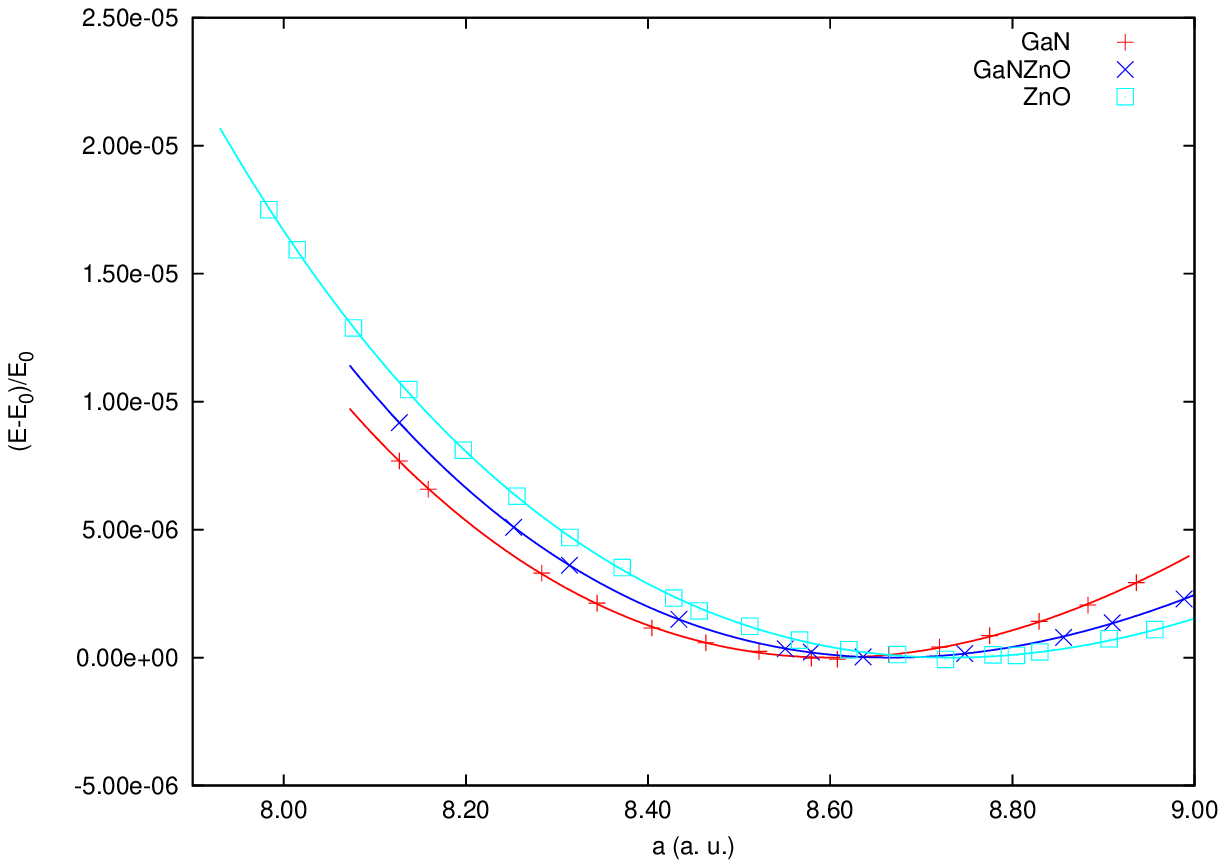}}
 \subfigure[]{\label{optim_PBE-PP} \includegraphics[width=7.00cm]{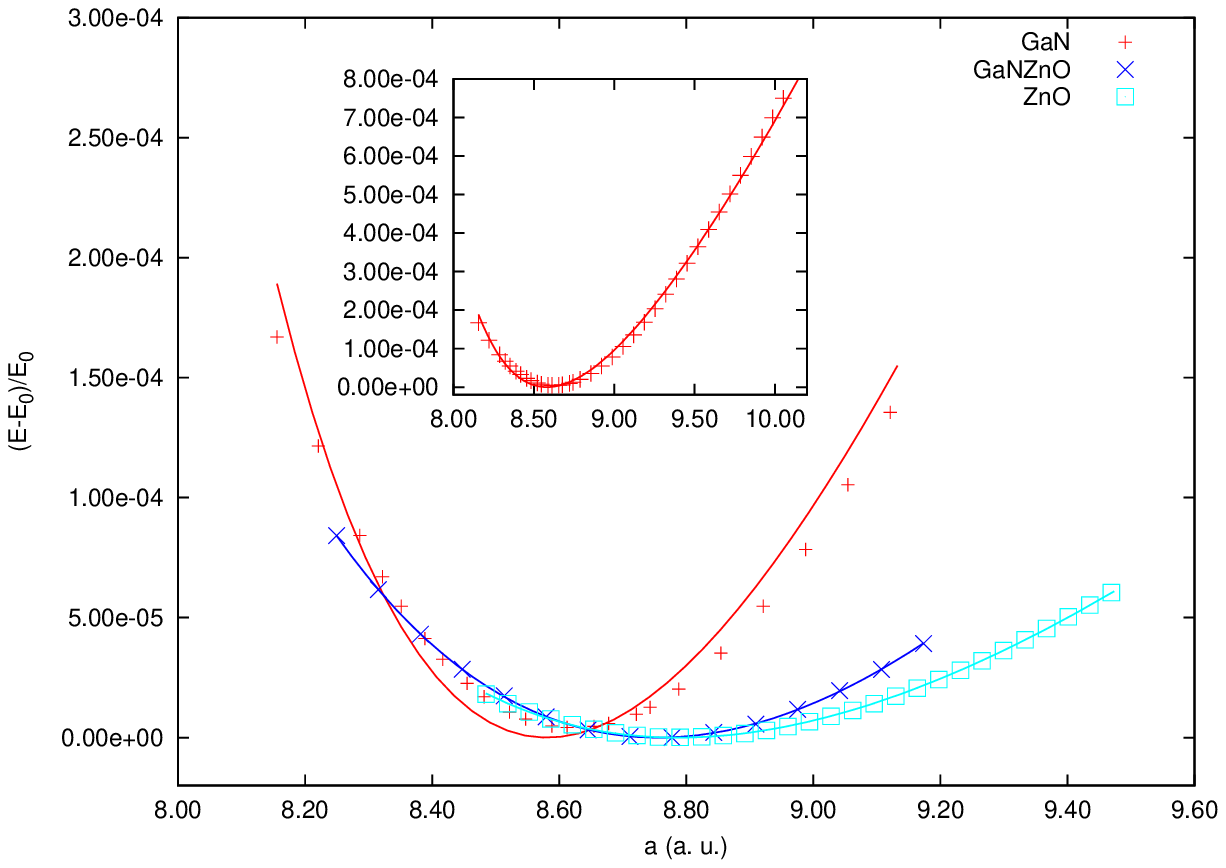}}
 \caption{Renormalized total energy vs lattice parameter of (GaN)$_{1}$/(ZnO)$_1$ superlattice and it parent binaries with: \subref{optim_LDA-LAPW} FP-LAPW(LDA), \subref{optim_PBE-LAPW} FP-LAPW(GGA) and \subref{optim_PBE-PP} LCPAO(GGA). Solid lines represent the fit by Murnaghan EOS.}
 \label{optim}
\end{figure}

Figure \ref{optim} shows the renormalized total energy of (GaN)$_1$/(ZnO)$_1$ superlattice performed over a set of different cell volumes. Equilibrium structural properties were obtained by fitting the calculated total energies versus volume data to the Murnaghan's equation of states \cite{murnaghan44} using both FP-LAPW and LCPAO methods within LDA and GGA approximations. The calculated equilibrium lattice parameter a, the bulk modulus B, its first pressure derivative B$’$ and the total energy E$_0$ are summarized in Table \ref{Table1} compared to their parent elements GaN and ZnO. No experimental data are available for the cubic (GaN)$_1$/(ZnO)$_1$ superlattice. The only experimental data are given by Maida et \textit{al.} for the (Ga$_{0.87}$Zn$_{0.13}$)(N$_{0.83}$O$_{0.16}$) ($a=b=3.189\mathring{A}$, $c=5.18367\mathring{A}$, $c/a=1.625$)\cite{maeda2005b} and the (Ga$_{0.93}$Zn$_{0.07}$)(N$_{0.90}$O$_{0.10}$) ($a=b=3.19\mathring{A}$, $c=5.1835\mathring{A}$, $c/a=1.625$)\cite{maeda2006b} in the wurtzite phase at the temperature of 299 K.

\begin{table}[!htp]
\caption{\label{Table1} Lattice constant a(\AA{}), bulk modulus B (GPa), pressure derivative of bulk modulus B$'$ and total energy E$_0$(eV) for (GaN)$_{1}$/(ZnO)$_1$ and their parents. FP-LAPW(LDA) and FP-LAPW(GGA) results are obtained by Wien2k \cite{blaha01} and LCPAO(GGA) by openmx codes \cite{openmx}.}
\renewcommand{\arraystretch}{1.4}
\begin{scriptsize}

\begin{tabular}{lllllll}
\hline\hline
\multicolumn{1}{l}{} &  & \multicolumn{3}{c}{Present work} & \multicolumn{1}{c}{} & \multicolumn{ 1}{c}{} \\ \cline{3-5}
\multicolumn{1}{l}{Composition} &  & FP-LAPW(LDA) & FP-LAPW(GGA) & LCPAO(GGA) & \multicolumn{1}{c}{Experiment} & \multicolumn{1}{c}{Other calculations} \\ 

\hline
\multicolumn{1}{l}{GaN} & \multicolumn{1}{l}{a} & 4.4607 & 4.5121 & 4.5406 & 4.52-4.55\cite{vanschilfgaarde97,bouarissa07,adachi05} &  4.46\cite{merad04} 4.50\cite{fonoberov03} 4.55\cite{riane09} 4.59\cite{bouarissa07}\\ 
\multicolumn{1}{l}{} & \multicolumn{1}{l}{} &  &  &  &  & 4.58\cite{sharma09} \\ 
\multicolumn{1}{l}{} & \multicolumn{1}{l}{B} & 199.5557 & 175.8859 & 193.3780 & 190\cite{serrano00,bouarissa07,sherwin91} & 202\cite{merad04} 176\cite{riane09} 165.59\cite{bouarissa07} \\  
\multicolumn{1}{c}{} & B$'$ & 3.9820 & 3.6525 & 7.4917 &  & 4.32\cite{merad04} 3.30\cite{riane09} 4.24\cite{bouarissa07} \\ 
\multicolumn{1}{c}{} & E$_0$ & -54307.19 & -54392.89 & -9471.06 &  &  \\ 

\hline
\multicolumn{1}{l}{(GaN)$_1$/(ZnO)$_1$} & a & 4.5859 & 4.5866 & 4.6332 &  &  \\ 
\multicolumn{1}{l}{} & B & 147.8437 & 145.5577 & 128.0100 &  &  \\ 
\multicolumn{1}{l}{} & B$'$ & 4.1421 & 4.3393 & 4.5090 &  &  \\ 
\multicolumn{1}{l}{} & E$_0$ & -210672.83 & -210632.00 & -17749.60 &  &  \\

\hline 
\multicolumn{1}{l}{ZnO} & \multicolumn{1}{l}{a } & 4.4938 & 4.6204 & 4.6497 & 4.47\cite{ashrafi07} & 4.634\cite{molepo11} 4.6329\cite{kalay09} 4.637\cite{uddin06}\\ 
\multicolumn{1}{l}{} & \multicolumn{1}{l}{} &  &  &  &  &  4.53\cite{dixit10} 4.489\cite{cui09} 4.337\cite{amrani07}\\ 
\multicolumn{1}{l}{} & \multicolumn{1}{l}{B} & 165.4759 & 129.7905 & 123.0910 &  & 124\cite{uddin06} 129.19\cite{molepo11} 165.9\cite{cui09} \\ 
\multicolumn{1}{l}{} & \multicolumn{1}{l}{} &  &  &  &  & 139.32\cite{kalay09} 166.649\cite{amrani07} \\ 
\multicolumn{1}{l}{} & B$'$ & 3.9386 & 4.0956 & 5.2790 &  & 4.33\cite{cui09} \\ 
\multicolumn{1}{l}{} & E$_0$ & -50838.65 & -50923.39 & -26028.51 &  &  \\ 
\hline\hline
\end{tabular}
\end{scriptsize}
\renewcommand{\arraystretch}{1}
\end{table}

\subsection{Electronic properties}
  \label{electronic}
\begin{figure}[!htp]
 \centering
 \subfigure[]{\label{GaNZnO-band-LDA-LAPW} \includegraphics[width=12.00cm]{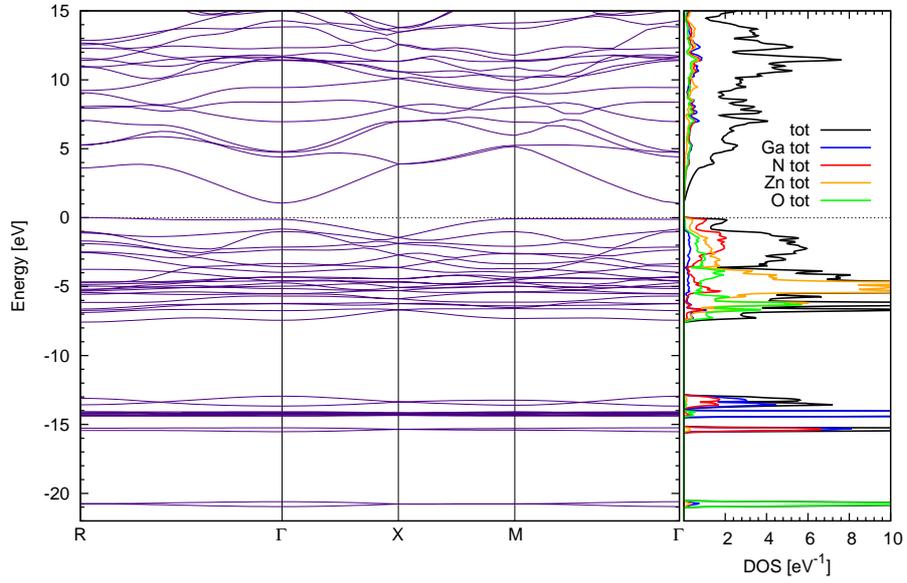}}
 \subfigure[]{\label{GaNZnO-band-LDA-PP} \includegraphics[width=12.00cm]{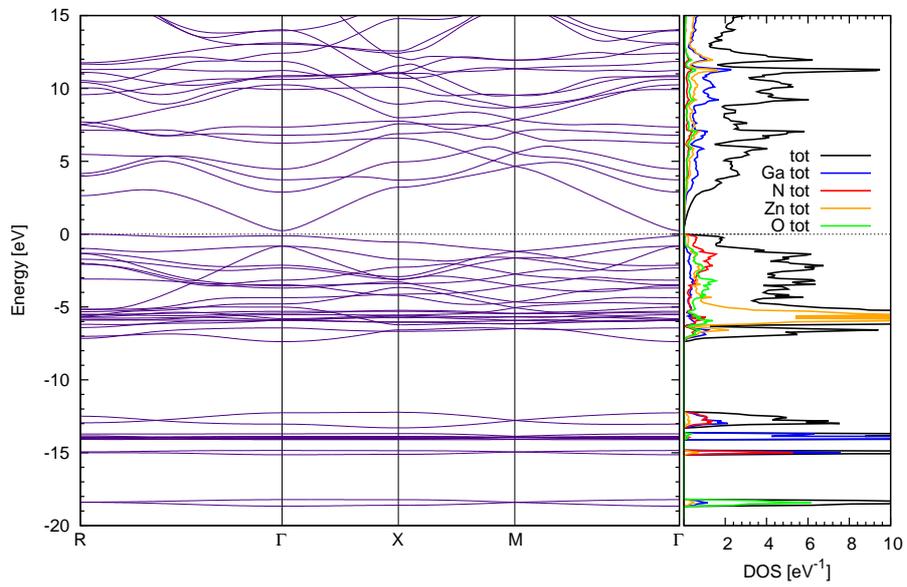}}
 \caption{Band structures and total DOS of (GaN)$_1$/(ZnO)$_1$ superlattice obtained by: \subref{GaNZnO-band-LDA-LAPW} mBJ-LDA and \subref{GaNZnO-band-LDA-PP} LCPAO(GGA).}
 \label{bandstructures}
\end{figure}

In order to give more accurate results for the superlattice, we have calculated the bandgap of GaN and ZnO compounds using both 
mBJ-LDA and mBJ-GGA. As illustrated in Table \ref{Table:Gap}, our bandgap calculations show that the mBJ-LDA is more improved 
than the mBJ-GGA in accordance with the results of Camargo-Martinez and Baquero \cite{camargo12} for GaN and ZnO. Figure 
\ref{bandstructures} illustrates the band structure of the (GaN)$_1$/(ZnO)$_1$ superlattice obtained by both FP-LAPW with 
mBJ-LDA (Figure \ref{GaNZnO-band-LDA-LAPW}) and LCPAO with GGA (Figure \ref{GaNZnO-band-LDA-PP}). By comparing to the binary 
parent elements GaN and ZnO, the bandgap of the heterostructure is lowered in amount considerably (see Table \ref{Table:Gap}).  
This finding is an indication of the strong positive bowing parameter for the GaNZnO quaternary alloys. S Wang and L-W Wang 
\cite{wang2010} show that ordering plays an important role in increasing the bandgap and reducing the bowing parameter for these 
alloys. As an indication, they obtained a value of 4.8 eV for wurtzite structure which is smaller than 11.5 eV \cite{huda08}. 
The bandgap of the (GaN)$_1$/(ZnO)$_1$ superlattice of 1.0419 eV is found to be slightly indirect at R point compared to the 
direct bandgap of 1.1710 eV. As consequence, from technological applications, the superlattice can be viewed then as 
approximately direct semiconductor and therefore to be considered as optically active material for photonic and optoelectronic 
devices.\\

\begin{figure}[!ht]
 \begin{center}
 \includegraphics[width=12.00cm]{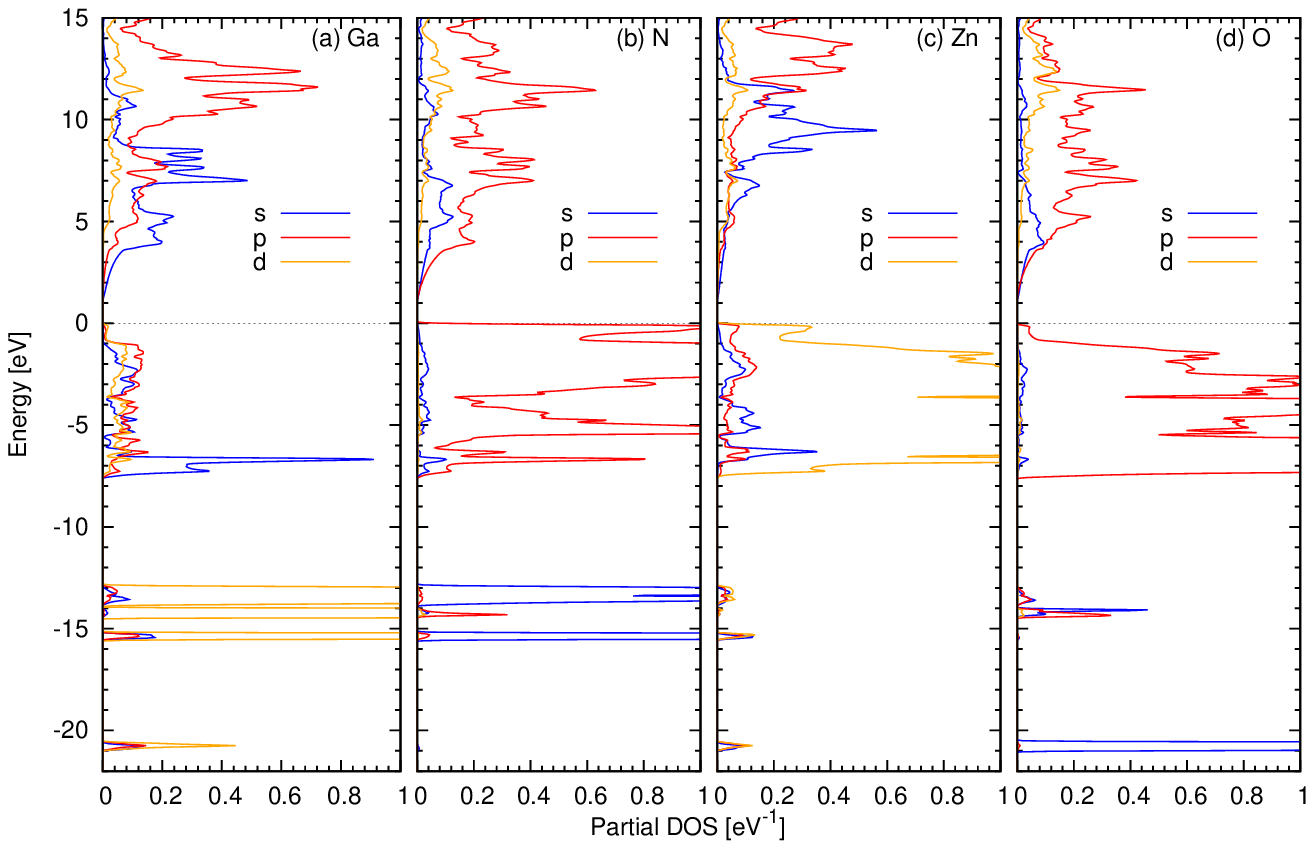}
 \caption{Partial DOS of (GaN)$_1$/(ZnO)$_1$ superlattice for : (a) Ga, (b) N, (c) Zn and (d) O calculated with mBJ-LDA.}
 \label{GaNZnO-pdos-LDA-LAPW}
 \end{center}
\end{figure}

The partial densities of states (PDOS) are illustrated in Figure \ref{GaNZnO-pdos-LDA-LAPW}. The lowest band for (GaN)$_{1}$/(ZnO)$_1$ superlattice at $\sim -21$ eV for mBJ-LDA (wien2k) and at $\sim -16$ eV for LCPAO(GGA) (openmx), is attributed essentially to the O $2s$ state and the following bands of the width 2.857 eV (mBJ-LDA) and 2.966 eV (LCPAO(GGA)) are mainly consisting of the localized Ga $3d$ and N $2s$ states. The upper valence band (UVB) of the width 7.683 eV (mBJ-LDA) and 7.535 eV (LCPAO(GGA)) is formed by the Ga $4s$, N $2p$, Zn $3d$ and O $2p$ states, but the maximum of the (UVB) is domination by the N $2p$ and the Zn $3d$ states in the Zn-N interface of our cubic superlattice. This gives rise to the $p-d$ repulsion which has been intensely debated as the responsible of the bandgap narrowing in the wurtzite like structure \cite{pan12}. In the other hand, the non negligible O $2p$ state in the (UVB) must be also considered as the origin of this visible light absorption mechanism produced by this bandgap narrowing. This was explained in the basis of the recent photoluminescence (PL) studies by some local inhomogeneity of the Zn and O atom densities leading to the impurity levels into the bandgap (ie empty impurity just above the valence band and/or filled impurity levels just below the conduction band) \cite{valentin10}. The conduction band originates from the contribution of the Ga, Zn, N and O atoms with variable amounts.\\

The effective mass of the (GaN)$_1$/(ZnO)$_1$ superlattice is an important property to be exploited in order to determine the efficiency of its PEC cell. In this context we have evaluated the effective masses of electron for the (GaN)$_1$/(ZnO)$_1$ superlattice as well as for the pure GaN and ZnO, from $\Gamma$ point along the principle directions of the Brillioun zone using the following expression:
\begin{equation*}
 E=\dfrac{\hbar^2 k^2}{2m^*}
\end{equation*}

Our results computed by mBJ-LDA are given in Table \ref{eff_mass_LAPW}, showing that the effective mass of electron is isotropic for the (GaN)$_1$/(ZnO)$_1$ superlattice and equal to $0.198 m_0$.
 
\begin{table}[hbp]
\caption{Calculated bandgaps by FP-LAPW(mBJ-LDA and mBJ-GGA) with mBJ obtained by Wien2k and LCPAO(GGA) by openmx. All energies are in eV.}
\renewcommand{\arraystretch}{1.4}
\begin{scriptsize}

\begin{tabular}{llccccll}
\hline\hline
\multicolumn{1}{l}{} & \multicolumn{1}{c}{} & \multicolumn{3}{c}{Present work} &  & \multicolumn{1}{c}{} & \multicolumn{1}{c}{} \\ \cline{3-5}
\multicolumn{1}{l}{Composition} & \multicolumn{1}{c}{} & mBJ-LDA & mBJ-GGA & LCPAO(GGA) & \multicolumn{1}{l}{Type} &  \multicolumn{1}{c}{Experiment} & \multicolumn{1}{c}{Other calculations} \\ \hline
\multicolumn{1}{l}{GaN} & E$_g(\Gamma^v \rightarrow \Gamma^c)$ & 3.1798 & 2.9524 & 1.6000 & D & 3.25 at RT \cite{novikov2008} & 1.930\cite{merad04} 1.747\cite{ahmed05} 1.811\cite{li11}\\ \multicolumn{1}{l}{} &  &  &  &  &  & 3.30 at $1.6 K$ \cite{novikov2008} &  1.520\cite{riane09} 1.706\cite{arbouche09} 3.13\cite{camargo12}\\ \hline
\multicolumn{1}{l}{(GaN)$_1$/(ZnO)$_1$} & E$_g(R^v \rightarrow \Gamma^c)$ & 1.0419 &  & 0.2409 & I &  & 3.360\cite{valentin10} (wurtzite phase) \\ 
\multicolumn{1}{l}{} & E$_g(\Gamma^v \rightarrow \Gamma^c)$ & 1.1710 &  & 0.3523 &  &  &  \\ \hline
\multicolumn{1}{l}{ZnO} & E$_g(\Gamma^v \rightarrow \Gamma^c)$ & 2.5935 & 1.2785 & 0.8037 & D & 3.27 at RT\cite{ashrafi07} &  0.710\cite{zhu08} 0.641\cite{karazhanov06} 0.695\cite{ulhaq12}\\ 
\multicolumn{1}{l}{} &  &  &  &  &  &  & 2.549\cite{ulhaq12} \\ 

\hline \hline
\end{tabular}

\label{Table:Gap}
\end{scriptsize}
\renewcommand{\arraystretch}{1}
\end{table}

\begin{figure}[h!]
 \begin{center}
  \includegraphics[width=12.00cm]{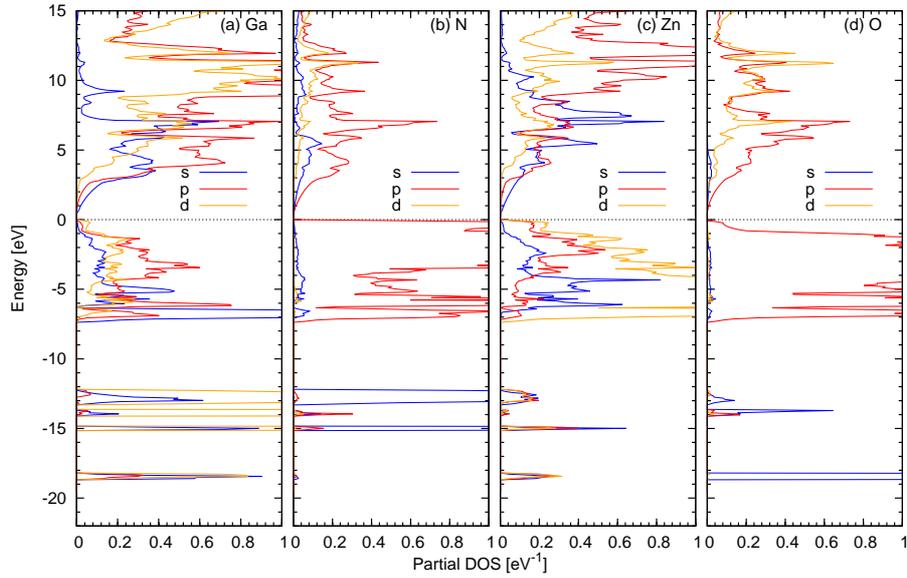}
 \caption{Partial DOS of (GaN)$_1$/(ZnO)$_1$ superlattice for : (a) Ga, (b) N, (c) Zn and (d) O calculated with LCPAO(GGA).}
 \label{GaNZnO-pdos-LDA-PP} 
 \end{center}
\end{figure}

\begin{table}[htbp]
\begin{center}
\caption{Calculated upper valence bandwith (UVBW) and total valence bandwith (TVBW). All energies are in eV. }
\begin{tabular}{lcccc}
\hline \hline
 & \multicolumn{2}{c}{UVBW} &  \multicolumn{2}{c}{TVBW}  \\ \cline{2-3} \cline{4-5}
\multicolumn{1}{l}{Composition} & mBJ-LDA &  LCPAO(GGA) & mBJ-LDA & LCPAO(GGA)\\ \hline
GaN & 6.7043 & 6.9739 & 16.6365 & 15.8717 \\ 
(GaN)$_1$/(ZnO)$_1$ & 7.6828 & 7.5351 & 21.0845 & 18.7575 \\ 
ZnO & 4.0370 & 6.0120 & 19.7857 & 17.7154 \\ 
\hline \hline
\end{tabular}
\end{center}
\label{UVBW-TVBW}

\end{table}

\begin{table}[htp]
 \begin{center}
  \caption{The calculated effective mass of electron along each directions from $\Gamma$ point obtained by mBJ-LDA.}
\begin{tabular}{lccc}
\hline \hline
 & \multicolumn{3}{c}{Effective mass (m$^*$/m$_0$)} \\ \cline{2-4}
  \multicolumn{1}{c}{Directions} & GaN & (GaN)$_1$/(ZnO)$_1$ & ZnO \\ \hline
  $\Gamma \rightarrow X$ & 0.147 & 0.199 & 0.190 \\ 
  $\Gamma \rightarrow L$ & 0.178 & \texttwelveudash & 0.197 \\ 
  $\Gamma \rightarrow R$ & \texttwelveudash & 0.181 & \texttwelveudash\\ 
  $\Gamma \rightarrow M$ & \texttwelveudash & 0.215 & \texttwelveudash \\ \hline
  Average & 0.162 & 0.198 & 0.194 \\ 
  Other Calc. & 0.165\cite{rodrigues98} & \texttwelveudash & 0.193\cite{karazhanov06b}\\
  \hline \hline
  \end{tabular}
  \label{eff_mass_LAPW} 
 \end{center}
\end{table}

\section{Conclusion}
\label{conclusion}
In this study we have investigated the structural and electronic properties of new structural cubic (GaN)$_1$/(ZnO)$_1$ superlattice. The structural optimisation as well as the electronic band structure are obtained using both full potential-linearized augmented plane wave (FP-LAPW) and linear combination of localized pseudo atomic orbital (LCPAO) methods. The bandgap of the parents elements GaN and ZnO are obtained to be more improved with mBJ-LDA than with mBJ-GGA. Our estimated mBJ-LDA bandgap value of the studied superlattice is slightly indirect (1.042 eV). This later is lowered considerably from those of the binary parent elements GaN and ZnO, indicating to the strong bowing parameter as is it observed in previous studies for the wurtzite structure. We have found that the origin of this decreasing is attributed to the $p-d$ repulsion of the Zn-N interface and the presence of the O $p$ electron. The electron effective mass is found to be isotropic. Good agreement is obtained between two used methods and with available theoretical and experimental data. We expected that our predictive results of the cubic (GaN)$_1$/(ZnO)$_1$ superlattice can be used as references for futur theoretical and experimental works. 

\section*{Acknowledgements}
This work is supported by the National Research Program (PNR) under project N\textdegree 43/08.

\end{document}